\documentclass[12pt]{article}
\usepackage[T2A]{fontenc}
  \usepackage[english]{babel}

\usepackage{latexsym}
\usepackage{amsmath}
\usepackage{amssymb}
\usepackage{bm}
\usepackage{graphicx}
\usepackage{indentfirst}
\usepackage{cite}

\newcommand{\be}{\begin{equation}}
\newcommand{\ee}{\end{equation}}

\newcommand{\bea}{\begin{eqnarray}}
\newcommand{\eea}{\end{eqnarray}}

\usepackage[dvips]{color}

\date{\today}

\title{On Wigner function of a vortex electron}

\author{Dmitry Karlovets}
\date{{\small Tomsk State University, Lenina 36, 634050 Tomsk, Russia}}

\begin{document}

\maketitle

\begin{abstract}
We derive a relatively simple and Lorentz-invariant expression for a Wigner function of a paraxial vortex electron described as a Laguerre-Gaussian wave packet.
\end{abstract}

While relativistic generalizations of a Wigner function \cite{W} of a quantum system have been shown to exist \cite{I1, I2, TMP, Ann}, 
these functions may still suffer from a lack of Lorentz invariance \cite{I2}. This is all the more so for wave packets with the non-Gaussian spatial profiles, 
such as a vortex state with an orbital angular momentum (OAM) $\ell$ \cite{Review}, an Airy beam \cite{Airy, Airy_El_Exp}, and their different generalizations. 
Although there are several works where the Wigner functions of such packets are derived (see, for instance, \cite{Gase, Simon, Singh, LG})
and even studied experimentally for twisted photons \cite{Singh, Twexp}, the problem of invariance of these functions under the Lorentz transformations for massive wave packets 
(first of all for vortex electrons, as they have recently been obtained experimentally \cite{Review}) remains unsolved.

One of the reasons why it has not been done for vortex electrons yet is that the very wave functions describing these states (the so-called Bessel beams and the Laguerre-Gaussian packets \cite{Review}) 
are usually written in terms of the non-invariant quantities and the transformation properties of these functions remain unclear. As noted by Bialynicki-Birula \cite{I2}, this lack of explicit invariance per se does not violate correct transformation properties of the observables. Nevertheless, aiming at relativistic applications in particle and hadronic physics \cite{JHEP}, it is desirable to know how the Wigner functions of the wave packets 
behave under the Lorentz boosts. To achieve this goal, the packets themselves need to be described in terms of the variables with definite transformation laws, 
which has been done for Gaussian packets of the massive neutrinos in \cite{Packets1, Packets2} and recently for vortex electrons in \cite{PRA}.

Here we derive a Wigner function of a massive particle with a phase vortex (say, the twisted electron) described via the paraxial Laguerre-Gaussian packet. The latter is Lorentz invariant for longitudinal boosts (along a propagation axis) and represents a massive generalization of the corresponding state of a twisted photon. Correct relativistic transformation of all variables was made possible by using an invariant approach to describe wave packets adopted from \cite{Packets1, Packets2}. As we argue in \cite{PRA}, it is convenient to choose the condition of paraxiality for massive particles to be invariant too, in contrast to the twisted photons. In this paper, we imply that the packet is wide in $x$-space compared to a Compton wavelength, $\sigma_{\perp} \gg \lambda_c = \hbar/mc \equiv 1/m$, or that it is narrow in $p$-space, $\sigma \sim 1/\sigma_{\perp} \ll m$. It is this Lorentz invariant condition of paraxiality that significantly simplifies the calculations and the resultant expression for the Wigner function. Indeed, as there is a small parameter in the problem now, it is enough to evaluate the corresponding integrals approximately, in a WKB fashion. Corrections to this approximation are mostly negligible for available beams, especially for small values of the OAM $\ell$ \cite{PRA}. 

By definition, the Wigner function of a scalar boson,
\begin{eqnarray}
& \displaystyle n({\bm r}, {\bm p}, t) = \int \frac{d^3 k}{(2\pi)^3}\,e^{i{\bm k}{\bm r}} \psi^*({\bm p} - {\bm k}/2, t)\, \psi({\bm p} + {\bm k}/2, t),
\label{ngen}
\end{eqnarray}
is \textit{not} manifestly Lorentz invariant. Here:
\begin{eqnarray}
& \displaystyle
\psi({\bm p}, t) = \psi({\bm p})\,e^{-it\varepsilon({\bm p})},\, \varepsilon({\bm p}) = \sqrt{{\bm p}^2 + m^2},\cr
& \displaystyle \int \frac{d^3p}{(2\pi)^3} |\psi({\bm p}, t)|^2 = \int \frac{d^3p}{(2\pi)^3} \frac{1}{2\varepsilon({\bm p})}\,|\Psi({\bm p}, t)|^2 = \int d^3 x\frac{d^3p}{(2\pi)^3}\, n ({\bm r}, {\bm p}, t) = 1 = \text{inv}.
\label{psip2}
\end{eqnarray}
Along with a ``non-relativistic'' wave function $\psi({\bm p})$, we have also introduced a ``relativistic'' one, 
$$
\Psi({\bm p}) = \sqrt{2\varepsilon({\bm p})}\,\psi({\bm p}).
$$

Within the paraxial regime, this function for a vortex boson in $(3+1)$D space-time represents an invariant Laguerre-Gaussian beam \cite{PRA},
\begin{eqnarray}
& \displaystyle \Psi_{\ell,n}^{\text{par}}({\bm p}) = \sqrt{\frac{n!}{(|\ell| + n)!}}\,\left(\frac{2\sqrt{\pi}}{\sigma}\right)^{3/2} \sqrt{2m}\, \left(\frac{p_{\perp}}{\sigma}\right)^{|\ell|} 
L_n^{|\ell|}(p_{\perp}^2/\sigma^2)\,\cr 
& \displaystyle\times \exp\left\{i\ell\phi_p - \frac{p_{\perp}^2}{2\sigma^2} - \frac{m^2}{\bar{\varepsilon}^2}\frac{(p_z - \bar{p})^2}{2\sigma^2}\right\},\
 \int\frac{d^3p}{(2\pi)^3}\,\frac{1}{2\varepsilon} |\Psi_{\ell,n}^{\text{par}}(p)|^2 = 1,\cr
& \displaystyle \Psi_{\ell,n}^{\text{par}}(\bm{r},t) = \int \frac{d^3p}{(2\pi)^3}\,\frac{1}{2\varepsilon}\, \Psi_{\ell,n}^{\text{par}}(\bm{p})\,e^{-ipx} = \sqrt{\frac{n!}{(|\ell| + n)!}}\, \frac{i^{2n + \ell}}{\pi^{3/4}\sqrt{2m}}\,\frac{(\rho/\sigma_{\perp}(t))^{|\ell|}}{\sigma_{\perp}^{3/2}(t)}\cr
& \displaystyle \times L_{n}^{|\ell|} \left(\rho^2/\sigma_{\perp}^2(t)\right) \exp\Big\{i\ell\phi_r - i\bar{p}_{\mu} x^{\mu} -i (2n + |\ell| + 3/2)\arctan(t/t_d) -\cr
& \displaystyle - \frac{1}{2\sigma_{\perp}^2(t)}\left(1 - i \frac{t}{t_d}\right)\left (\rho^2 + \bar{\varepsilon}^2(z-\bar{u}t)^2/m^2\right)\Big\},\
\int d^3 x\,2\bar{\varepsilon}\, |\Psi_{\ell,n}^{\text{par}}(\bm{r},t)|^2 = 1,
\label{psip}
\end{eqnarray}
where
\begin{eqnarray}
& \displaystyle
t_d = \frac{\bar{\varepsilon}}{\sigma^2},\ \sigma_{\perp}(t) = \sigma^{-1}\sqrt{1 + (t/t_d)^2},\ \bar{u} = \bar{p}/\bar{\varepsilon},\cr
& \displaystyle {\bm r} = \{\rho \cos\phi_r, \rho \sin\phi_r, z\},\ {\bm p} = \{p_{\perp} \cos\phi_p, p_{\perp} \sin\phi_p, p_z\}.
\label{var}
\end{eqnarray}
Here, $n + 1 = 1,2,3,..$ defines a number of radial maxima of the probability distribution, the term with $\arctan(t/t_d)$ represents a so-called Gouy phase, which is also Lorentz invariant for a massive particle, 
$\bar{\varepsilon} = \sqrt{\bar{\bm p}^2 + m^2}$ is a mean energy of the packet (note that within the paraxial approximation we deal with the positive-energy solutions only), $\bar{\bm p} = \{0,0,\bar{p}\}$ its mean momentum, $\bar{p}_{\mu} x^{\mu} = \bar{\varepsilon} t - \bar{p}z$, and in the paraxial regime one can always put $\varepsilon \equiv \varepsilon({\bm p}) \approx \bar{\varepsilon}$ in the pre-exponential factor.

The function $ \Psi_{\ell,n}^{\text{par}}(\bm{r},t)$ represents an approximate paraxial solution to the Klein-Gordon equation, applicable when the momentum uncertainty is small, $\sigma \ll m$.
Both the wave functions in Eq.(\ref{psip}) are the eigenfunctions of an angular momentum operator $\hat{L}_z$, which is either $\hat{L}_z = -i\partial/\partial \phi_r$ or $\hat{L}_z = -i\partial/\partial \phi_p$, 
depending on the representation. Unlike the Bessel beam \cite{Review}, the Laguerre-Gaussian state (\ref{psip}) is well-localized, which is somewhat even more important for massive particles than for photons.
Indeed, a wave packet of an electron cannot be focused to a spot with a size $\sigma_{\perp}\sim 1/\sigma$ smaller than the Compton wave length $\lambda_c$ without creating electron-positron pairs \cite{BLP}.
While the model of the Bessel beam does not take this circumstance into account (its transverse momentum $\kappa$, unlike $\sigma$, can be arbitrarily large), 
the Laguerre-Gaussian packets themselves are applicable only in the one-particle (paraxial) regime with $\sigma_{\perp} \gg \lambda_c$, thus representing the positive-energy states.

Finally, it is worth noting that a non-relativistic Laguerre-Gaussian wave function,
\begin{eqnarray}
& \displaystyle
\psi_{\ell,n}({\bm r},t) = \sqrt{2m}\,\Psi_{\ell,n}^{\text{par}}({\bm r},t)\Big|_{\bar{p}\ll m}e^{i m t},
\label{nonrel}
\end{eqnarray}
obeys the Schr\"odinger equation \textit{exactly}. As the overall phase $e^{i m t}$ does not contribute to the Wigner function, the resultant function that we obtain hereafter
also describes a non-relativistic vortex electron.

Now we evaluate the integral over ${\bm k}$ in (\ref{ngen}) in the paraxial (akin to WKB) approximation, that is, we neglect the $\mathcal O(k^2)$ terms, evaluate the resultant Gaussian integrals, 
and arrive at the following Wigner function 
of a vortex scalar:
\begin{eqnarray}
& \displaystyle n^{\text{par}}_{n,\ell} ({\bm r}, {\bm p}, t) = 8 \frac{n!}{(n + |\ell|)!}\,\left(\frac{p_{\perp}}{\sigma}\right)^{2|\ell|} \left[L_{n}^{|\ell|}(p_{\perp}^2/\sigma^2)\right]^2
\cr
& \displaystyle \times \exp\Big\{-\sigma^2\left(\rho^2 + \bar{\varepsilon}^2(z-ut)^2/m^2\right) - \frac{1}{\sigma^2} \left(p_{\perp}^2 + m^2 (p_z - \bar{p})^2/\bar{\varepsilon}^2\right)\Big\},
\label{npar}
\end{eqnarray}
which is a Lorentz scalar for longitudinal boosts because 
$$
\rho^2 + \bar{\varepsilon}^2(z-ut)^2/m^2 = \text{inv},\ p_{\perp}^2 + m^2 (p_z - \bar{p})^2/\bar{\varepsilon}^2 = \text{inv},
$$ 
and where $u = {\bm p}/\varepsilon({\bm p})$. 
The correction to this expression due to the phase $\exp\{i\ell\phi_p\}$ is $\mathcal O(\ell\sigma^2)$ and, therefore, 
it must be neglected in the paraxial regime.

This Wigner function has the following properties:
\begin{eqnarray}
& \displaystyle \int d^3x\, n^{\text{par}}_{n,\ell} ({\bm r}, {\bm p}, t) = \frac{1}{2\bar{\varepsilon}}\, |\Psi_{n,\ell}^{\text{par}}({\bm p})|^2,\
\int d^3 x\frac{d^3p}{(2\pi)^3}\, n^{\text{par}}_{n,\ell} ({\bm r}, {\bm p}, t) = 1.
\label{nparp}
\end{eqnarray}
On the other hand,
\begin{eqnarray}
& \displaystyle \int \frac{d^3p}{(2\pi)^3}\, n^{\text{par}}_{n,\ell} ({\bm r}, {\bm p}, 0) = \frac{\bar{\varepsilon}}{m} \left(\frac{\sigma}{\sqrt{\pi}}\right)^3 \exp\left\{-\sigma^2 \left(\rho^2 + \bar{\varepsilon}^2 z^2/m^2\right)\right\} \ne 2\bar{\varepsilon}\,|\Psi_{n,\ell}^{\text{par}} ({\bm r}, 0)|^2
\label{nparx}
\end{eqnarray}
is independent of $n$ and $\ell$ and it does not have the typical doughnut-like spatial structure.

However, as follows from the exponent in (\ref{npar}), the following estimate also takes place in the pre-exponential factor in the paraxial regime:
\begin{eqnarray}
& \displaystyle
\frac{p_{\perp}}{\sigma} \approx \sigma \rho.
\label{eq}
\end{eqnarray}
This yields an alternative representation for the Wigner function,
\begin{eqnarray}
& \displaystyle 
n^{\text{par}}_{n,\ell} ({\bm r}, {\bm p}, t) = 8 \frac{n!}{(n + |\ell|)!}\,\left(\sigma\rho\right)^{2|\ell|} \left[L_{n}^{|\ell|}(\sigma^2\rho^2)\right]^2
\cr
& \displaystyle \times \exp\Big\{-\sigma^2\left(\rho^2 + \bar{\varepsilon}^2(z-ut)^2/m^2\right) - \frac{1}{\sigma^2} \left(p_{\perp}^2 + m^2 (p_z - \bar{p})^2/\bar{\varepsilon}^2\right)\Big\},
\label{nparalt}
\end{eqnarray}
with the following properties:
\begin{eqnarray}
& \displaystyle \int \frac{d^3p}{(2\pi)^3}\, n^{\text{par}}_{n,\ell} ({\bm r}, {\bm p}, 0) = 2\bar{\varepsilon}\,|\Psi_{n,\ell}^{\text{par}} ({\bm r}, 0)|^2,\ 
\text{but}\ \int d^3x\, n^{\text{par}}_{n,\ell} ({\bm r}, {\bm p}, t) \ne \frac{1}{2\bar{\varepsilon}}\, |\Psi_{n,\ell}^{\text{par}}({\bm p})|^2.
\label{nparalt2}
\end{eqnarray}
In other words, such a function predicts the correct doughnut-like spatial profile in $x$-space but fails to do so in $p$-space.

One could arrive at the function (\ref{nparalt}) in a more rigorous way by starting from the representation of the Wigner function in (\ref{ngen}) 
in terms of the wave functions $\psi(\bm{r},t)$ in $x$-space instead. Hence, there are two alternative expressions for the Wigner function of a vortex spinless boson, 
which are completely equivalent (that is, interchangeable) within the paraxial regime. Both these functions are everywhere positive and can be called quasi-classical.
If needed, one can rewrite the pre-exponential factors of these Wigner functions in an equivalent $x-p$ symmetric form (akin to \cite{Simon}) by using the following paraxial equalities:
\begin{eqnarray}
& \displaystyle 
\left(\sigma\rho\right)^{2|\ell|} \left[L_{n}^{|\ell|}(\sigma^2\rho^2)\right]^2 \approx \left(\rho p_{\perp}\right)^{|\ell|} \left[L_{n}^{|\ell|}(\rho p_{\perp})\right]^2 \approx
\left(\rho p_{\perp}\right)^{|\ell|} \left[L_{n}^{|\ell|}\left(\frac{1}{2}\left(\sigma^2\rho^2 + \frac{p_{\perp}^2}{\sigma^2}\right)\right)\right]^2.
\label{nparsymm}
\end{eqnarray}
Therefore, this factor does not effectively depend upon the packet's width $\sigma$ at all. Beyond the paraxial approximation, this is no longer valid of course.

The estimate (\ref{eq}) also allows one to make quantitative predictions even without the calculation of observables. In particular, as follows from Eqs.(\ref{npar}),(\ref{nparalt}),
$$
\left(\sigma\rho\right)^{2|\ell|} \approx \left(\rho p_{\perp}\right)^{|\ell|} \Rightarrow \langle \rho \rangle \langle p_{\perp }\rangle \sim |\ell|
$$
when $|\ell| \gg n$ and where $\langle \rho \rangle$ is a mean value of the beam radius and $\langle p_{\perp }\rangle$ is a mean absolute value of the transverse momentum (see \cite{PRA}).

Let us now turn to a fermion with spin. Following the standard approach to generalize the scalar Wigner function (see, for instance, \cite{I1, I2}), one can define the fermionic function as follows\footnote{There are two definitions used in literature: one with a Dirac conjugate, $\bar{\Psi}$, and another with a Hermitian conjugate, $\Psi^{\dagger}$. While in (\ref{nfermex}) we stick to the former definition,
its alternative would be $n_f({\bm r}, {\bm p}, t) =\int\frac{d^3 k}{(2\pi)^3}\, \frac{1}{\sqrt{2\varepsilon({\bm p} - {\bm k}/2)2\varepsilon({\bm p} + {\bm k}/2)}}\,e^{i{\bm k}{\bm r}} \Psi^{\dagger}_f({\bm p} - {\bm k}/2, t) \Psi_f({\bm p} + {\bm k}/2, t)$. In the paraxial regime with the positive-energy states, the two expressions coincide.}:
\begin{eqnarray}
& \displaystyle n_f({\bm r}, {\bm p}, t) =\frac{1}{2m}\int\frac{d^3 k}{(2\pi)^3}\, e^{i{\bm k}{\bm r}} \bar{\Psi}_f({\bm p} - {\bm k}/2, t) \Psi_f({\bm p} + {\bm k}/2, t).
\label{nfermex}
\end{eqnarray}
where the factor $1/2m$ provides the correct normalization, 
$$
\int d^3x \frac{d^3p}{(2\pi)^3}\, n_f ({\bm r}, {\bm p}, t) = 1 = \text{inv}.
$$
Here, also
$$
\Psi_f({\bm p}, t) = \frac{u({\bm p})}{\sqrt{2\varepsilon({\bm p})}}\, \Psi({\bm p}, t),\ \int\frac{d^3 p}{(2\pi)^3}\frac{1}{2\varepsilon({\bm p})}\,\Psi_f^{\dagger}({\bm p}, t)\Psi_f({\bm p}, t) = 1 = \text{inv},
$$
where $\Psi({\bm p}, t)$ is the scalar wave function used above, and $u({\bm p})$ is a bispinor normalized as $\bar{u}({\bm p}) u({\bm p}) = 2m,\,u^{\dagger}({\bm p}) u({\bm p}) = 2\varepsilon({\bm p}), \bar{u} = u^{\dagger}\gamma^0$.

Next, in the paraxial approximation we have
\begin{eqnarray}
& \displaystyle \frac{1}{2m} \frac{\bar{u}({\bm p}-{\bm k}/2) u({\bm p}+{\bm k}/2)}{\sqrt{2\varepsilon({\bm p}-{\bm k}/2)2\varepsilon({\bm p}+{\bm k}/2)}} = \frac{1}{2\varepsilon({\bm p})} + \mathcal O(k^2) 
= \frac{1}{2\bar{\varepsilon}} + \mathcal O(k^2, ({\bm p} - \bar{\bm p})^2).
\label{uparax}
\end{eqnarray}
As a result, the spin does not make a contribution and the paraxial Wigner function of a fermion coincides with that of a boson:
\begin{eqnarray}
& \displaystyle n_f^{\text{par}} ({\bm r}, {\bm p}, t) = n^{\text{par}} ({\bm r}, {\bm p}, t)
\label{uparax2}
\end{eqnarray}
with $n^{\text{par}} ({\bm r}, {\bm p}, t)$ from Eq.(\ref{ngen}), (\ref{npar}), or (\ref{nparalt}).

Finally, note that the Wigner functions derived cannot be directly applied to the massless case and compared with the corresponding functions of the twisted photons. 
Indeed, the paraxial approximation implies that it is the packet's width $\sigma \ll m$ that is the smallest parameter of the problem, not $m$.
The formal limit of $m \rightarrow 0$ means that the particle's Compton wave length tends to infinity and the invariant condition of paraxiality $1/\sigma \gg \lambda_c = 1/m$ loses its sense.
If needed, one can derive \textit{a non-paraxial} expression for the Wigner function by using the corresponding wave functions from \cite{PRA}. 
Such a Wigner function can be made finite in the massless limit, analogously to Ref.\cite{Ann} where a simplest example of this kind has been found. 

We would like to emphasize, however, that for available beams of electrons the non-paraxial effects do not exceed the relative values of $10^{-4}-10^{-3}$ \cite{PRA}. 
Therefore, the relatively simple paraxial expressions for the Wigner function derived in this paper must suffice for physical applications 
in the overwhelming majority of practical cases. One of such potential applications is a study of scattering problems with the twisted wave packets in which Wigner functions of the incoming states 
naturally arise in the theoretical formalism \cite{JHEP}. 

\

This work was conducted within a government task of the Ministry of Education and Science of the Russian Federation, project No.\,$3.9594.2017/8.9$.

\end{document}